\documentstyle[prl,aps]{revtex}
\input epsf

\begin{document}

\title{Inflation without slow roll}
\author{Thibault Damour}
\address{Institut des Hautes Etudes Scientifiques, 91440
Bures-sur-Yvette, France and \\
DARC, CNRS-Observatoire de Paris, 92195 Meudon, France}
\author{Viatcheslav F. Mukhanov}
\address{Theoretische Physik, Ludwig-Maximilians-Universit\"at, \\
Theresienstr. 37, D-80333 M\"unchen, Germany}
\maketitle

\begin{abstract}
We draw attention to the possibility that inflation (i.e. accelerated
expansion) might continue after the end of slow roll, during a period of
fast oscillations of the inflaton field $\varphi $. This phenomenon takes place
when a mild non-convexity inequality is satisfied by the potential $V(\varphi 
)$. The presence of such a period of $\varphi $-oscillation-driven
inflation can substantially modify reheating scenarios.
 In some models the effect of
these fast oscillations might be imprinted on the primordial perturbation
spectrum at cosmological scales.
\end{abstract}

\vglue 1cm

The inflationary paradigm has become a widely accepted
element of early universe cosmology \cite{G81,Lbook}. This paradigm offers the
attractive possibility of resolving many of the shortcomings of standard hot
big bang cosmology whilst providing an explanation for the origin of
structure in the universe \cite{MC,GP,Hawking,Starobinskii,BST}. Although
the underlying physical ideas of  inflation seem well established,
 which concrete inflationary scenario is realized in the very
early universe is unknown. There exist, at present,  many inflationary 
scenarios,
which, despite some common features, differ greatly in their details.
The crucial ingredient of nearly all known successful inflationary scenarios
is a period of ``slow roll'' evolution of the inflaton field, during which a
quasi-homogeneous scalar field $\varphi $ changes very slowly, so that
its kinetic energy $\dot{\varphi}^{2}/2$ during inflation remains always
much smaller than its potential energy $V(\varphi )$ \cite{Lchao,Lnew,ASnew}. 
Such a period of $\varphi $-domination is well known to generate an
accelerated expansion of the universe, thereby providing a natural mechanism
for solving the causality and homogeneity puzzles of hot big bang cosmology.
The standard inflationary lore assumes that the end of the slow roll
evolution marks the end of inflation, and that it is followed by a
non-inflationary period during which the inflaton $\varphi $ oscillates
rapidly around the minimum of its potential $V(\varphi )$.

The main aim of this work is to draw attention to the possibility that
inflation might continue after the end of slow roll, during a period of fast
oscillations of $\varphi $. Such a period of $\varphi $-oscillation-driven
inflation presents novel physical characteristics which can be 
crucial for the theory of reheating and which may modify some features of the
fluctuation spectra expected from inflation. The possibility of 
$\varphi$-oscillation-driven inflation has (as far as we know) not been 
previously
noticed.  Most authors  work mainly with the
simplest renormalizable tree-level potentials, like $V(\varphi )=\frac{1}{2}
\,m^{2}\,\varphi ^{2}$ or $V(\varphi )=\frac{1}{4}\,\lambda \,\varphi ^{4}$.
For such potentials, and more generally for convex functions $V(\varphi )$
(with vanishing minimum), the end of slow roll necessarily marks the end of
inflation. By contrast, we shall see that
 {\it non convex} functions $V(\varphi ) $ can,
 as long as a certain inequality is satisfied, entail the continuation
of inflation after the end of slow roll. 
Such non convex potentials might arise in various ways.
Let us only mention two possibilities. 
First, supergravity and/or superstring physics may generate very general
types of non-renormalizable potentials depending on several scalar fields 
$\phi_{i}$, $V(\phi _{i})=\Lambda ^{4}\,{\cal V}\,(\phi 
_{i}/\widetilde{m}_{P})$. The inflaton $\varphi $ would then correspond to a 
relatively flat direction in the space of scalar fields. 
Second, loop-contributions to a classically $\varphi$-independent
potential naturally generate a {\it logarithmic} potential
for large values of $\varphi $ in some supersymmetric models
 \cite{W81,DR83,dss,lr}:
 $V(\varphi )\sim A\,\ln (\varphi /\mu )+B$. When the
coefficient $A$ is positive (which is the case of the models of Refs. 
\cite{W81,DR83}, except when the gauge coupling contribution dominates
the scalar couplings one), such a logarithmic potential is not convex 
($V_{,\varphi \varphi }<0$).
However, it is not clear whether such loop-corrected potentials can
sustain the type of oscillatory inflation discussed below because,
when $\varphi$ becomes small the fields whose masses depend on $\varphi$ 
may become light or tachyonic, so that one must consider a multi-field 
dynamics.

In units where $\widetilde{m}_{P}=(4\pi G)^{-1/2}=1$, a conveniently
redundant set of evolution equations for a scalar-driven (flat) Friedmann
cosmology read 
\begin{equation}
\ddot{\varphi}+3H\dot{\varphi}+V_{,\varphi }=0 \, , \label{eq1.1}
\end{equation}
\begin{equation}
H^{2}=\frac{2}{3}\,\epsilon \, , \label{eq1.2}
\end{equation}
\begin{equation}
\dot{\epsilon}=-3H(\epsilon +p)=-3H\,\dot{\varphi}^{2} \, , \label{eq1.3}
\end{equation}
\begin{equation}
\frac{\ddot{a}}{a}=-\frac{1}{3}\,(\epsilon +3p) \, .\label{eq1.4}
\end{equation}
Here, $H\equiv \dot{a}/a$ is the physical-time expansion rate, 
$V_{,\varphi }\equiv \partial V/\partial \varphi $, $\epsilon \equiv 
\frac{1}{2}\,\dot{\varphi}^{2}+V(\varphi )$ denotes the energy density of the 
scalar field, and $p\equiv \frac{1}{2}\,\dot{\varphi}^{2}-V(\varphi )$ its 
pressure. Only
two equations among Eqs.~(\ref{eq1.1})-(\ref{eq1.4}) are independent. The ``slow 
roll'' regime is the case where one can neglect $\frac{1}{2}\,\dot{\varphi}^{2}$ 
in $\epsilon$ and $\ddot{\varphi}$ in Eq.~(\ref{eq1.1}). It is easy to see that 
slow roll can take place only when the potential $V\left( \varphi \right) $  
satisfies the following conditions: $\varepsilon _{1}\equiv 
\dot{\varphi}^{2}/2V\simeq \frac{1}{12}\,(V_{,\varphi }/V)^{2}\ll 1$ and 
$\varepsilon _{2}\equiv \ddot{\varphi}/3H\dot{\varphi}\simeq 
\frac{1}{6}\,V_{,\varphi \varphi }/V - \varepsilon_1 \ll 1$.
Then the effective adiabatic index of the scalar matter, $\gamma _{{\rm
slow\,roll}}\equiv (\epsilon +p)/\epsilon =\dot{\varphi}^{2}/\epsilon \simeq
2\,\varepsilon _{1}$ is much smaller than one. This guarantees that the
right-hand side of Eq.~(\ref{eq1.4}) is positive, i.e. that the expansion is
accelerated.

The slow roll conditions are sufficient, but {\it not necessary} to maintain
inflation. We derive below the general conditions on the potential $V\left(
\varphi \right) $ under which inflation can proceed even during a
stage of fast oscillations of the scalar field. For simplicity, we consider
an even potential, $V(\varphi )=V(-\varphi )$, which has its minimum at 
$\varphi =0$. Slow roll is then followed by a stage where $\varphi $
oscillates symmetrically around 0. For generic potentials (as we shall check
below), the $\varphi $-oscillations become ``adiabatic'' soon after the exit
from slow roll (i.e. when $|\varphi | \lesssim 1$), in the sense that the 
expansion rate $H$ becomes much smaller than the  frequency of oscillations 
$\omega $. In the adiabatic approximation $H\ll \omega $, one can find 
approximate solutions of Eqs.~(\ref{eq1.1})-(\ref{eq1.4}) by separating 
the two time scales characterizing the evolution \cite{Turner} (we have
also checked numerically the validity of this approximation).
 On the fast, oscillation 
time scale, one first neglects the Hubble damping terms $\propto 3H$ in 
Eqs.~(\ref{eq1.1}) and (\ref{eq1.3}), and gets $\varphi$
as a function of time by inverting the integral obtained by writing the
conservation of energy, $\epsilon \simeq {\rm const}=V_{m}$, where $V_{m}\equiv 
V(\varphi _{m})$ denotes the maximum current value of the
potential energy: 
\begin{equation}
t-t_{0}=\pm \int d\varphi \,[2(V_{m}-V(\varphi ))]^{-1/2} \, . \label{eq1.5}
\end{equation}
The full oscillation period is 
\begin{equation}
T\equiv 2\pi /\omega =4\int_{0}^{\varphi _{m}}d\varphi \,[2(V_{m}-V(\varphi
))]^{-1/2} \, . \label{eq1.6}
\end{equation}

\begin{figure}
\begin{center}
\leavevmode\epsfbox{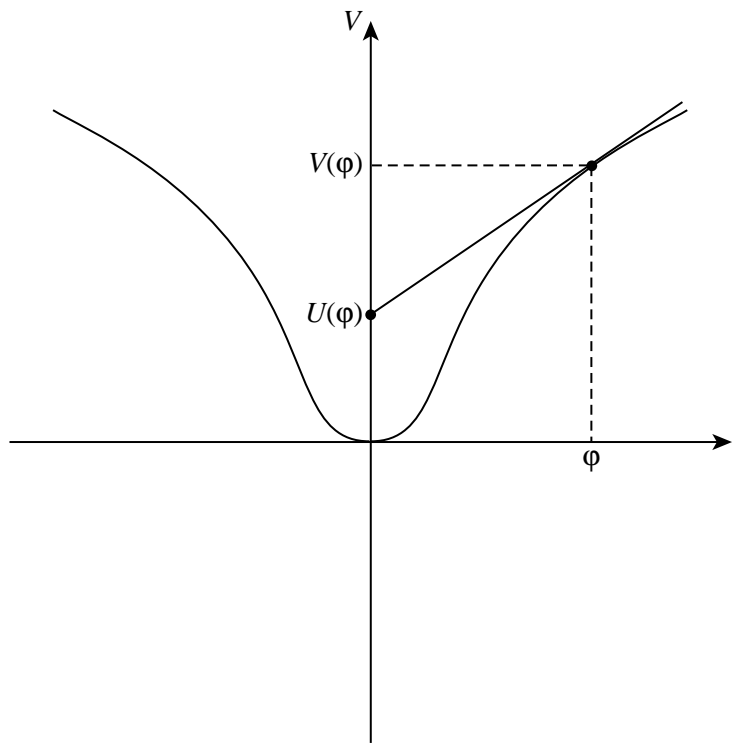}
\end{center}
\vskip 1pc
\caption{Intercept $U(\varphi)$ of the tangent to the curve $V
= V(\varphi)$ at the point $\varphi$. Inflation continues after slow roll if
the time-average of $U(\varphi (t))$ over one $\varphi$-oscillation is
positive.}
\label{fig1}
\end{figure}

On the longer, expansion time scale, the energy $\epsilon $ is slowly
drained out by the Hubble damping terms. From Eq.~(\ref{eq1.3}) the
oscillation-averaged fractional energy loss reads 
\begin{equation}
\langle \dot{\epsilon}/\epsilon \rangle =-3\langle H(\epsilon +p)/\epsilon
\rangle \simeq -3H\gamma \, , \label{eq1.7}
\end{equation}
where the angular brackets denote a time average, and where $\gamma \equiv
\langle (\epsilon +p)/\epsilon \rangle \equiv \langle 
\dot{\varphi}^{2}/\epsilon \rangle $ is the average adiabatic index of scalar 
matter (the averaged equation of state is $p=(\gamma -1)\,\epsilon $). From 
Eq.~(\ref{eq1.4}), one sees that the condition for inflation to continue during 
the $\varphi$-oscillation regime (i.e. the condition for accelerated expansion 
$\ddot{a}>0$) is $\gamma <\frac{2}{3}$. This condition can be rewritten in 
several ways in terms of the potential $V(\varphi )$. Indeed, neglecting the 
expansion of the universe, one easily derives the following set of equalities
\begin{equation}
\gamma =\frac{\langle \dot{\varphi}^{2}\rangle }{\epsilon }=\frac{\langle
\varphi \,V_{,\varphi }\rangle }{V_{m}}=2\left( 1-\frac{\langle 
V\rangle}{V_{m}}\right) 
=2\,\frac{\int_{0}^{1}d\widehat{\varphi}\,(1-\widehat{V}
)^{1/2}}{\int_{0}^{1}d\widehat{\varphi }\,(1-\widehat{V})^{-1/2}} \, , 
\label{eq1.8}
\end{equation}
where in the last one $\widehat{\varphi }\equiv \varphi /\varphi _{m}$ and 
$\widehat{V}\equiv V(\varphi )/V_{m}$. Using Eqs.~(\ref{eq1.8}), the condition 
$\gamma < \frac{2}{3}$ for having a $\varphi$-oscillation-driven inflation can 
also be written as 
\begin{equation}
\langle V - \varphi \, V_{,\varphi} \rangle > 0 \, , \label{eq1.9}
\end{equation}
which has a very simple geometrical interpretation. Indeed, the quantity 
$U(\varphi )\equiv V(\varphi )-\varphi \,V_{,\varphi }\,(\varphi )$ is simply
the ``intercept'' of the tangent to the curve $V=V(\varphi )$ at the point 
$\varphi$, i.e. its intersection with the vertical axis $\varphi =0$ (see
Fig.~\ref{fig1}). The condition is therefore that the time-average, over an
oscillation, of the intercept $\langle U\rangle
=T^{-1}\int_{0}^{T}dt\,U(\varphi (t))$ must be positive. In the case where 
$V_{0}=V(\varphi =0)$ is either strictly zero (as in Fig.~\ref{fig1}) or very 
small compared to $V_{m}$, this geometrical interpretation shows that one needs
potentials $V(\varphi )$ which are sufficiently non convex near the maximum
amplitude $\varphi _{m}$ to compensate the negative $U$'s contributed by the
convex part of $V(\varphi )$ near the bottom $\varphi =0$. Then inflation
continues as long as $\varphi _{m}$ is larger than some value $\varphi _{c}$
defining the size of the convex core (around $\varphi =0$) of $V(\varphi )$.

Let us apply our general considerations to a simple class of 
potentials within which the condition (\ref{eq1.9}) can be satisfied, without 
fine tuning, in many models. Namely, we consider potentials having at most
polynomial growth when $\varphi \gg \varphi _{c}:V(\varphi )\sim \varphi ^{q}
$, with any (positive) real exponent $q$. In these models, slow roll takes 
place when $\varphi \gg 1$
and terminates around $\varphi \sim 1$.
When $\varphi \ll 1$ one can use the adiabatic oscillation approximation
because the ``adiabaticity parameter'' is seen, using equations given
above, to be generically $ H/ \omega \sim \varphi$.
 To be more concrete let us take, for instance, the class of models, 
\begin{equation}
V(\varphi )=\frac{A}{q}\,\left[ \left( \frac{\varphi ^{2}}{\varphi _{c}^{2}}
+1\right) ^{\frac{1}{2}q}-1\right] \, , \label{eq1.10}
\end{equation}
containing one dimensionless real parameter $q>0$, and two dimensionful
parameters: an overall scale $A\sim ({\rm mass})^{4}$, and the scale $\varphi 
_{c}$ (which could be the weak scale) determining the size of the convex core 
of $V(\varphi )$. In
the limit $q\rightarrow 0$, this gives a logarithmic potential $V(\varphi )=
\frac{1}{2}\,A\,\ln (1+\varphi ^{2}/\varphi _{c}^{2})$ similar to the ones
naturally arising in the supersymmetric models mentioned above. If $\varphi
_{c}\ll 1$ (in Planck units), then, after the end of slow roll, one can have
many oscillations with $\varphi _{m}\gg \varphi _{c}$. At  $\varphi \gg
\varphi _{c}$ the potential $V\left( \varphi \right) $ can be well
approximated by the power law potential $V(\varphi )\simeq A\,q^{-1}(\varphi
/\varphi _{c})^{q}$, or a logarithmic one $V(\varphi )\simeq A\,\ln (\varphi
/\varphi _{c})$ when $q\rightarrow 0$.

For such power law potentials one easily obtains from Eqs.~(\ref{eq1.8})  the 
average adiabatic index, 
\begin{equation}
\gamma =\frac{2q}{q+2} \, , \label{eq1.11}
\end{equation}
as well as the adiabatic evolution laws for the various relevant physical
quantities, 
\begin{mathletters}
\label{eq1.12}
\begin{eqnarray}
a\propto t^{\frac{2}{3\gamma }}=t^{\frac{q+2}{3q}} \, ,
\label{eq1.12a} \\
\epsilon =V(\varphi _{m})\propto t^{-2}\propto a^{-\frac{6q}{q+2}} \, ,
\label{eq1.12b} \\
\varphi _{m}\propto t^{-\frac{2}{q}}\propto a^{-\frac{6}{q+2}} \, ,
\label{eq1.12c} \\
V_{,\varphi \varphi }\,(\varphi _{m})\propto t^{\frac{2(2-q)}{q}}\propto
a^{6\,\frac{2-q}{2+q}} \, .
\label{eq1.12d}
\end{eqnarray}
\end{mathletters}

We note that: (i) inflation continues during the $\varphi$-oscillation-driven 
expansion if $\gamma <2/3$, i.e. $q<1$; (ii) the
logarithmic case, $V\simeq A\,\ln (\varphi /\varphi _{c})$, interestingly
leads to quasi-exponential inflation. [When $q\rightarrow 0$ Eqs.~(\ref{eq1.11}) 
and (\ref{eq1.12}) get modified because $\gamma $ is not zero but only 
logarithmically small, $\gamma =(\ln \,(\varphi _{m}/\varphi _{c}))^{-1}$, 
leading, e.g., to $\epsilon =\epsilon _{0}-3A\,\ln (a/a_{0})$, $\varphi 
_{m}\propto a^{-3}$ and $a(t)\propto \exp \left[ -\frac{1}{2}\,A(t_{{\rm 
end}}-t)^{2}\right] $]; (iii) the total number of $e$-folds $N$ of this new type 
of inflation is determined (from Eq.~(\ref{eq1.12c}) by the hierarchy between 
$\varphi _{{\rm initial}}\sim \widetilde{m}_{P}$ (end of slow roll) and $\varphi 
_{{\rm final}}\sim \varphi _{c}$ (convex core of $V(\varphi )$): 
\begin{equation}
N\simeq \frac{q+2}{6}\ \ln \,\left( \frac{\widetilde{m}_{P}}{\varphi _{c}}
\right) \, . \label{eq1.13}
\end{equation}

Let us also mention that the above statements can (with some changes) 
 be extended to the case of negative powers 
$-2 < q < 0$ in Eq.~(\ref{eq1.10}). This corresponds to a potential $V 
(\varphi)$ which climbs up to a constant when $\vert \varphi \vert \gg 
\varphi_c$ and defines a trough for $\vert \varphi \vert < \varphi_c$. 
In this case, slow roll ends when  $\vert \varphi \vert < \varphi_{\rm slow \; 
roll} \sim \varphi_c^{-\frac{q}{2 - q}} \ll 1$. The field $\varphi$ oscillates 
rapidly (on the Hubble time scale) when $ \varphi_c \ll \varphi_m
\ll \varphi_{\rm slow \; roll}$. Eq.~(\ref{eq1.11}) does not apply; $\gamma$ is 
power-law small, and a quasi-exponential inflation takes place
 during $\varphi$-oscillations.

The qualitative explanation of why, despite the fast oscillations of the scalar
field, we can still have an accelerated expansion of the universe is
simple. When the potential $V\left( \varphi \right) $ satisfies
the condition (\ref{eq1.9}), the scalar field spends a dominant fraction
of each period of oscillation on the upper parts of the potential, where the
kinetic energy $\dot{\varphi}^{2}/2$ is small compared to $V\left( \varphi
\right)$. Therefore, the main contribution to the averaged effective
equation of state comes from $V\left( \varphi \right) .$

The classical estimate Eq.~(\ref{eq1.13}) severely constrains the total number 
of 
$e$-folds spent during any $\varphi -$oscillation-driven inflation. In the case 
of a logarithmic potential ($q\simeq 0$), and the extreme case of a weak-scale
core $\varphi _{c}\sim 300\,{\rm Ge}V$, one gets $N\simeq 12$. This is
sizable, but still small compared to the total needed duration of inflation, 
$N_{{\rm tot}}>65$. Moreover, if we impose some other reasonable restrictions
on the model then the number of allowed $e$-folds can become even smaller. In
particular, if one requires that the observable cosmological perturbations
were produced during the slow roll stage of the evolution of the field $\varphi$
and that the mass of $\varphi$ near the core is smaller than the Planck
mass, then $N\lesssim 5 (2+q) / (2-q)$. However, even in such a case, the 
$\varphi$-oscillation-driven inflation can still lead to some interesting 
consequences
which we briefly discuss below.

Up to this point we have discussed the evolution of a classical homogeneous
background $\varphi (t)$, neglecting the backreaction of the quantum
fluctuations of the inflaton field. However, these fluctuations can be strongly 
amplified because of the fast oscillations of $\varphi$ and can have a dramatic 
backreaction effect on the evolution
of the background. Let us consider the fluctuations
of the gauge-invariant variable $v=a\,[\delta \varphi ^{(gi)}+(\dot{\varphi}
/H)\,\Phi ]$ \cite{MFB} which describes the coupled scalar-matter gravity
fluctuations. The mode of $v$ with comoving spatial momentum $k$ satisfies
the equation (in conformal time: $v^{\prime \prime }\equiv \partial
^{2}v/\partial \eta ^{2}$) 
\begin{equation}
v^{\prime \prime }+(k^{2}-U_0(\eta ))\,v=0 \, , \label{eq1.14}
\end{equation}
\begin{equation}
U_0=z^{\prime \prime }/z=a^{2}\,[-V_{,\varphi \varphi }-4V_{,\varphi 
}\,\dot{\varphi}/H-7\dot{\varphi}^{2}+2H^{2}+2\dot{\varphi}^{4}/H^{2}] \, , 
\label{eq1.15}
\end{equation}
where $z\equiv a\,\dot{\varphi}/H$. The effective potential for scalar
fluctuations $U_0(\eta )$ exhibits novel features during the $\varphi $
oscillations associated with a non convex potential $V(\varphi )$. Indeed,
the dominant term in Eq.~(\ref{eq1.15}) is proportional to the squared 
``effective'' mass of the scalar field $m_{\varphi }^{2}\equiv \,V_{,\varphi 
\varphi }$ which is mostly {\it \ negative}  and oscillates with an increasing 
frequency and an increasing amplitude (see Eq.~(\ref{eq1.12d})).
 The resulting, mostly {\it positive}, oscillations of the effective 
potential $U_0(\eta) \simeq 
-a^2 \, V_{,\varphi \varphi}$ are much more efficient at amplifying the 
fluctuations of $v$ than even the broadly resonant Mathieu-equation-type ones 
recently discussed \cite{KLS}. We shall leave to future work a discussion
 of the
effects of such a new type of super-broad resonance and only note here that
the associated fast exponential growth of scalar fluctuations will quickly
modify the classical, adiabatic evolution presented above, and can bring
interesting new features in reheating theory. One of the simplest examples
of such new features is the possibility, due to the {\it increase} of the
oscillation frequency $\omega \sim (-V_{,\varphi \varphi }\,(\varphi
_{m}))^{1/2}$ as $\varphi _{m}$ sinks down, to generate more and more
massive particles coupled to $\varphi $, thereby alleviating the usual
obstacles \cite{KLR96} to producing the superheavy grand-unified-theory
(GUT) bosons needed in GUT baryogenesis scenarios.

This very effective parametric amplification of cosmological perturbations
puts also specific imprints  on the primordial perturbation spectrum. 
However, if
the presently discussed mechanism terminates inflation, then, because of the
above mentioned limits on the duration of the fast-oscillation inflationary
stage, it can  only influence the fluctuation spectra on small length scales.
Nevertheless, it is easy to imagine how (with some amount of fine tuning)
one can translate the effect of our mechanism on cosmologically relevant
length scales. It suffices to consider hybrid-type inflationary models where
a first bout of (slow roll, plus oscillation-driven) inflation linked to the
evolution of $\varphi $ is followed by a secondary bout of inflation driven
by another scalar field. With adequate tuning of the duration of the
secondary inflation the effects of the $\varphi $-oscillation-driven
inflation might be imprinted on cosmological scales.

\smallskip

We thank Andrei Linde and Antonio Riotto for useful discussions.

\newpage 

\end{document}